\documentclass[aps,prd,onecolumn,eqsecnum,amsmath,nofootinbib,preprintnumbers]{revtex4}%
\newcounter{example}[section]

\usepackage{amsmath}
\usepackage{color,graphicx,float,subfigure}
\usepackage{amsfonts,amssymb,theorem,mathrsfs,times}
\usepackage{bm}
\usepackage{amsmath}
{\theorembodyfont{\upshape}
	}
{\theorembodyfont{\upshape}
	}
{\theorembodyfont{\upshape}
	}
{\theorembodyfont{\upshape}
	}
{\theorembodyfont{\upshape}
	}
{\theorembodyfont{\upshape}
	}

\newcommand{\dalm}{\kern1pt\vbox{\hrule height 0.9pt\hbox{\vrule width
			0.9pt\hskip 2.5pt\vbox{\vskip 5.5pt}\hskip 3pt\vrule width
			0.3pt}\hrule height 0.3pt}\kern1pt}

\begin{document}
	\title{A unified topological classification of circular orbits for charged particles in black hole spacetimes}
	
	%
	
\author{Yong Song$^{a}$\footnote{e-mail
		address: syong@cdut.edu.cn}}
\author{Jia Li$^{a}$\footnote{e-mail 
		address: lijia@cdut.edu.cn (corresponding author)}}
\author{Yiting Cen$^{a}$\footnote{e-mail
		address: 2199882193@qq.com}}
\author{Kai Diao$^{b}$\footnote{e-mail
		address: kai\_diao@qq.com}}
	\author{Xiaofeng Zhao$^{a}$\footnote{e-mail
		address: 59403804@qq.com}}
\author{Shunping Shi$^{a}$\footnote{e-mail
		address: shishunping13@cdut.edu.cn}}

	
	\affiliation{$^a$
	College of Physics, Chengdu University of Technology, Chengdu, Sichuan 610059,
	China,\\$^b$ Yin bin Campus, Chengdu Technological University Yinbin, 644000 Sichuan, China}

	\date{\today}
	
	\begin{abstract}
	 The study of circular orbits offers profound insights into the structure of spacetime around black holes. While the topological properties of these orbits are well-established for neutral particles, the influence of electric charge—particularly for massless particles—remains a subject of exploration. In this work, we employ a topological current $\phi$-mapping approach to systematically investigate the circular orbits of charged test particles in static, spherically symmetric black hole spacetimes with flat, anti-de Sitter (AdS), and de Sitter (dS) asymptotics. We demonstrate that the particle's charge significantly alters the topological classification of both timelike and null circular orbits. A key finding is that for multi-horizon black holes, if a circular orbit with fixed angular momentum and charge exists between two neighboring horizons, there will always be at least one unstable null and one unstable timelike circular orbit. Outside the outermost horizon, the asymptotic behavior of spacetime and the specific charge ratio crucially determine the topological charge $W$, dictating the existence and stability of orbits. Our results, validated through Reissner–Nordström (RN), RN-AdS, and RN-dS examples, extend the topological orbit classification framework and provide a foundation for potential applications in environments where effective charge dynamics may be relevant, such as magnetized plasmas around black holes.
	\end{abstract}
	

	\maketitle


\section{Introduction}
In recent years, the study of circular orbits in black hole spacetimes has provided profound insights into gravitational physics and related astrophysical phenomena~\cite{Khan:2019gco,Khan:2020sya,Jumaniyozov:2024hlg,Khan:2024jez,Jumaniyozov:2025uwo,Nishonov:2025pgq}. These orbits are essential for understanding the strong-field regime of gravity, black hole accretion processes, and the emission characteristics of active galactic nuclei and X-ray binaries. Topological methods have emerged as a powerful tool to analyze the existence and stability of circular orbits in a coordinate-invariant manner, offering a new perspective on the global structure of black hole spacetimes.

In 2017, Cunha, Berti, and Herdeiro introduced a novel topological approach, which, without relying on specific field equations, studies the photon sphere (PS) or light ring (LR) stability for ultra-compact objects~\cite{Cunha:2017qtt}. Using the Brouwer degree of a continuous map, they found that the LRs of compact objects always come in pairs. Though this result is generally valid, an important exception arises when degenerate light rings are present~\cite{Hod:2017zpi}. Further, Cunha and Herdeiro advanced the study by generalizing it to a four-dimensional stationary, axisymmetric, asymptotically flat black hole spacetime~\cite{Cunha:2020azh}. They proved that at least one standard LR (i.e., an unstable LR) exists outside the black hole horizon for each rotation sense by calculating the winding number of the vector field defined by an effective potential on the orthogonal $(r,\theta)$-space. Moreover, following their topological argument, horizonless ultra-compact objects like boson stars exhibit an even number of non-degenerate LRs. Currently, there are already many studies based on their work~\cite{Ghosh:2021txu,Wei:2020rbh}. Wei, based on Duan’s topological current $\phi$-mapping theory~\cite{Duan:1979ucg,Duan:1984ws} and the work of Cunha et al., computed the topological charge for a black hole with asymptotically flat, AdS, and dS boundary behaviors~\cite{Wei:2020rbh}. They found that all these black holes admit a minus one topological charge, indicating the existence of at least one standard PS. They also found that Duan’s topological current $\phi$-mapping can be used for the study of black hole thermodynamics, providing new insights into this field~\cite{Wei:2021vdx,Wei:2022dzw}.

Previous studies have focused on null circular orbits, while the case for timelike circular orbits (TCOs) is a bit more complicated. There have been several studies on the topological properties of TCOs~\cite{Wei:2022mzv,Ye:2023gmk,Shahzad:2024oxb}. The main conclusions are that, for a generic black hole spacetime, TCOs with fixed angular momentum always come in pairs, with one being stable and the other being unstable. A positive or negative winding number corresponds to a stable or unstable TCO, respectively. This result is universal and is independent of the angular momentum of the particle and the black hole parameters. 

However, the aforementioned studies primarily concern neutral test particles. In realistic astrophysical environments, such as accretion disks and jets, plasma effects and electromagnetic fields are ubiquitous. Charged particles moving in the magnetized vicinity of black holes can exhibit rich dynamics \cite{Ruffini:1994,Rohrlich:2007,Lim:2015oha,Lei:2021koj,Pugliese:2011py,Kurbonov:2023uyr}. Understanding the orbital topology of charged particles is thus a natural and necessary extension. Furthermore, while the concept of massless charged particles is speculative within the Standard Model \cite{Case:1962zz,Weinberg:1980kq}, they arise in certain theoretical frameworks \cite{Azzurli:2014lha,Fairoos:2017lnm,Lechner:2014kua} and can serve as useful proxies for studying the influence of charge in the extreme limit where rest mass becomes negligible. The dynamics of such particles are also required by specific solutions in general relativity, like the charged Vaidya metric \cite{Chatterjee:2015cyv}.

In 2024, Xu and Wei investigated the topological phenomena of TCOs for charged test particles in asymptotically flat black holes and discovered that the charge-to-mass ratio (specific charge) affects the topological behavior of TCOs \cite{Ye:2024sus}. Inspired by their work, we further explore the topological properties of circular orbits for charged particles, including massless charged particles, across a wider range of spacetime backgrounds.

This work aims to address the following questions: How does the charge of a particle influence the topological classification of both timelike and null circular orbits? How do different asymptotic behaviors (flat, AdS, dS) modify this classification? Furthermore, what are the topological properties of orbits residing between multiple horizons of a black hole? By applying the $\phi$-mapping topological current method to charged particles in static, spherically symmetric black holes with various asymptotics, we provide a unified topological classification. Our findings reveal that charge introduces new topological phases and that the inter-horizon region exhibits a universal topological invariant. Although the net charge of astrophysical black holes is likely small, our model could be relevant in effective descriptions of magnetized plasmas where particles exhibit effective charge-to-mass ratios. This study extends the topological orbit classification paradigm and could guide future study involving more realistic, rotating black holes, where such topology might be imprinted in observational signatures like quasi-periodic oscillations (QPOs) or shadow images. In X-ray binaries, quasi-periodic oscillations (QPOs) are thought to originate from orbital motion and resonance mechanisms in the inner accretion disk, where the fundamental frequencies—such as the orbital, radial, and vertical epicyclic frequencies—are tied to the stability of circular orbits~\cite{Abramowicz:2001bi,Motta:2013wwa}. The topological charge $W$ introduced in our study serves as a global indicator of this orbital stability. Consequently, a topological transition (e.g., a change in W) in the orbit structure, potentially induced by the effective charge of plasma constituents or spacetime asymptotics, could imprint detectable signatures in QPO spectra. Similarly, the black hole shadow, as observed by the Event Horizon Telescope, is directly shaped by the photon sphere, which comprises unstable null circular orbits~\cite{EventHorizonTelescope:2019dse}. The topological number $W$ associated with these null orbits dictates their existence and stability. Therefore, changes in this topological invariant due to charge or background curvature could, in principle, alter the shadow's characteristics, such as its size and subring structure. Thus, our topological framework not only advances theoretical understanding but also establishes a potential link between fundamental spacetime topology and electromagnetic observables, bridging the gap between abstract gravitational concepts and astrophysical phenomenology.

The present paper is organized as follows: In Sec.~\ref{section2}, we will review the charged black hole and the topological approach for circular orbits. In Sec.~\ref{section3}, we will study the topology of circular orbits of charged particles between two neighboring horizons. In Sec.~\ref{section4}, we will study the topology of circular orbits for charged particles in asymptotically flat, AdS, and dS black holes. In Sec.~\ref{section5}, we summarize and discuss our results. In Appendix.~\ref{Appendix}, we present the results on the topological properties of circular orbits in the case of $V=V_-$.

\section{Topological approach for circular orbits}\label{section2}
In this work, we only consider the static spherically symmetric black hole, and its line element in standard spherical coordinates $(x=t,r,\theta,\phi)$ can be assumed as~\cite{Wei:2020rbh}
\begin{align}
\label{metric}
ds^2=-f(r)dt^2+\frac{1}{g(r)}dr^2+h(r)(d\theta^2+\sin^2\theta d\phi^2)\;,
\end{align}
where $f(r)$, $g(r)$ and $h(r)$ are functions that depend only on the radial coordinate $r$, and the horizon is defined by $f(r_h)=0$. A Maxwell field 
\begin{align}
A=A_\mu dx^\mu, \quad\mathrm{and} \quad A_\mu=(A_t(r),0,0,0)\;.
\end{align}
is assumed to exist, and $A_\mu$ are the components of the electromagnetic $4-$potential. At infinity, we assume $A_t(r)$ has the following asymptotic behavior
\begin{align}
A_t(r\to\infty)\sim -\frac{Q}{r}\;.
\end{align}
where $Q$ is the charge of the black hole. The Lagrangian for the motion of a charged particle in different references has different forms~\cite{Ruffini:1994,Rohrlich:2007,Kurbonov:2023uyr,Ye:2024sus}. Here, we generalize the form in Ref.~\cite{Ye:2024sus} based on the results of Ref.~\cite{Ori:1991} and write the Lagrangian as
\begin{align}
\label{Lagrangian1}
\mathcal{L}&=\frac{1}{2}g_{\alpha\beta}\dot{x}^\alpha\dot{x}^\beta+q A_\alpha \dot{x}^\alpha\;,\\
\label{Lagrangian}
&=\frac{1}{2}\bigg[-f\dot{t}^2+\frac{\dot{r}^2}{g}+h(\dot{\theta}^2+\sin^2\theta\dot{\phi}^2)\bigg]+q A_t\dot{t}\;,
\end{align}
where the dot represents differentiation with respect to the proper time, and $q$ represents the charge per unit mass of a massive charged particle or the charge of a massless charged particle. The four-velocity $(\dot{x}^\alpha)$ satisfies
\begin{align}
\label{normalized}
	g_{\alpha\beta}\dot{x}^\alpha\dot{x}^\beta=-\mu^2\;,
\end{align}
where $\mu^2=1, 0$ correspond to the timelike (massive) particles and null (massless) particles, respectively.

The equations of motion for the charged particle can be derived from Eq. (\ref{Lagrangian1}) using the Euler–Lagrange equation, which are consistent with the results in~\cite{Ori:1991}, and can be written as follows:
\begin{align}
\dot{x}^\alpha\nabla_\alpha \dot{x}^\beta=q F^\beta{}_\gamma \dot{x}^\gamma\;,
\end{align}
where $F_{\alpha\beta}\equiv A_{\alpha,\beta}-A_{\beta,\alpha}$.

Since the Lagrangian (\ref{Lagrangian}) does not depend explicitly on the variables $t$ and $\phi$, the following two conserved quantities exist
\begin{align}
\label{E}
&p_t\equiv\frac{\partial\mathcal{L}}{\partial\dot{t}}=-\bigg(f\dot{t}-q A_t\bigg)=-E\;,\\
\label{L}
&p_\phi\equiv\frac{\partial\mathcal{L}}{\partial\dot{\phi}}=h\sin^2\theta\dot{\phi}=L\;,
\end{align}
where $E$ and $L$ are the energy and angular momentum of the charged particles, respectively. These conserved quantities arise from the symmetry of the Lagrangian under translations in $t$ and $\phi$, and hold everywhere in the spacetime, independent of the particle's trajectory. In the spherically symmetric case, $L$ can always be chosen to be greater than zero, i.e., $L>0$. 

Considering the metric (\ref{metric}) and (\ref{normalized}), one can obtain
\begin{align}
\label{motion}
-f\dot{t}^2+\frac{\dot{r}^2}{g}+h\dot{\theta^2}+h\sin^2\theta\dot{\phi}^2+\mu^2=0\;.
\end{align}
To study the circular orbits in the spacetime, one can review the effective potential method in~\cite{deFelice:1968um,Bardeen:1973tla,Wilkins:1972rs}. Here, we follow the Ref.~\cite{Cunha:2017qtt,Wei:2022mzv}, and define the kinetic term $\mathcal{K}$ and the potential term $\mathcal{V}$ as
\begin{align}
\label{K}
&\mathcal{K}=\frac{\dot{r}^2}{g}+h\dot{\theta^2}\;,\\
\label{V0}
&\mathcal{V}=-f\dot{t}^2+h\sin^2\theta\dot{\phi}^2+\mu^2\;.
\end{align}
Therefore, the motion of the charged particle (\ref{motion}) becomes
\begin{align}
\mathcal{K}+\mathcal{V}=0\;.
\end{align}
Note that the kinetic term $\mathcal{K}\ge 0$, and the inequality is only saturated when $\dot{r}=\dot{\theta}=0$. The motion of the particles can be completely governed by the effective potential $\mathcal{V}$. 

For a neutral particle, when studying the LR, Cunha et al.~\cite{Cunha:2017qtt,Cunha:2020azh} introduced the regular potential functions $H_{\pm}(r,\theta)$, which are independent of the orbital parameters $E$ and $L$. However, in the case of TCOs, it is not possible to introduce a potential function that is independent of the orbital parameters. Nevertheless, it is possible to study the topological properties of circular orbits with fixed angular momentum, as shown in~\cite{Wei:2022mzv}. For a charged particle, both $L$ and $q$ will affect their circular orbits. Similarly, we cannot introduce a regular potential function that is independent of the orbital parameters, even in the massless case. However, following the approach in Ref.~\cite{Wei:2022mzv}, we can investigate the topological properties of circular orbits for charged particles with fixed angular momentum and charge.

Considering Eqs.(\ref{E}) and (\ref{L}), the potential (\ref{V0}) can be reduced to
\begin{align}
	\mathcal{V}=-\frac{1}{f}(E-V_+)(E-V_-)\;,
\end{align}
where
\begin{align}
	\label{V}
	V_\pm=-q A_t\pm \sqrt{f\bigg(\mu^2+\frac{L^2}{h\sin^2\theta}\bigg)}\;,
\end{align}
The circular motion of the charged particles can be described using the following equations:
\begin{align}
\label{circularorbit}
	\partial_\mu V_{\pm}=0,\quad V_{\pm}=E\;.
\end{align}
It should be noted that $\partial_\mu V_{\pm}$ here not only depends on $r$ and the angular momentum $L$, but also on the charge $q$ of the charged particle. 

In order to give a global topology, we require that the values of the angular momentum and the charge do not change the asymptotic behavior of $\partial_r V_{\pm}$ at the boundary of the $(r,\theta)$ plane. Here we only consider the case of $V=V_+$. For $V=V_-$, one can see the results in Appendix~\ref{Appendix}. Following Ref. \cite{Wei:2020rbh}, we deﬁne a vector field $\phi=(\phi^r,\phi^\theta)$
\begin{align}
	\label{phi}
	\phi^r=\frac{\partial_r V}{\sqrt{g_{rr}}},\quad\phi^\theta=\frac{\partial_\theta V}{\sqrt{g_{\theta\theta}}}\;,
\end{align}
in a ﬂat vector space. For a massless charged particle, we have $\mu^2=0$. From Eqs.(\ref{V}) and (\ref{phi}), one can get
\begin{align}
	\label{phir0}
	&\phi^r=-q A_t'\sqrt{g}+\frac{L(hf'-fh')}{2h^{3/2}\sin\theta}\sqrt{\frac{g}{f}}\;,\\
	\label{phit0}
	&\phi^\theta=-\frac{\sqrt{f}L\cos\theta}{h\sin^2\theta}\;.
\end{align}
For a massive charged particles, we have $\mu^2=1$. Then, one obtains
\begin{align}
	\label{phir}
	&\phi^r=-q A_t'\sqrt{g}+\frac{f'h^2\sin^2\theta+L^2(hf'-fh')}{2h^{3/2}\sin\theta\sqrt{h\sin^2\theta+L^2}}\sqrt{\frac{g}{f}}\;,\\
	\label{phit}
	&\phi^\theta=-\frac{\sqrt{f}L^2\cos\theta}{h\sin^2\theta\sqrt{h\sin^2\theta+L^2}}\;.
\end{align}

It follows that $\partial^\mu V_\pm\partial_\mu V_\pm=(\phi^r)^2+(\phi^\theta)^2=||\phi||^2$, where $||\phi||=\sqrt{\phi^a\phi_a}$. Hence, in terms of the vector ﬁeld $\phi^a$, $\phi^a=0\Leftrightarrow ||\phi||=0$. Obviously, $\phi^a=0$ corresponds to the circular orbits of charged particles and $\theta=\pi/2$ as expected. Deﬁne an angle $\Omega$ such that $\phi^r=\phi\cos\Omega$, $\phi^\theta=\phi\sin\Omega$, and $\Omega=\arctan(\phi^\theta/\phi^r)$.

\begin{figure}[H]
	\centering
	\includegraphics[width=3in]{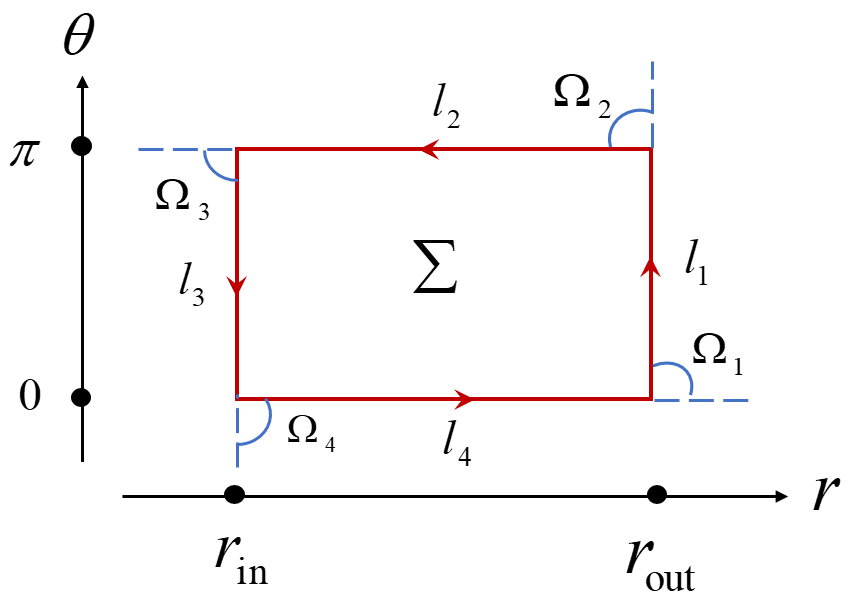}
	\caption{Representation of the contour $C=\sum_i\cup l_i$ (which encloses $\Sigma$) on the $(r,\theta)$ plane. The curve $C$ has a positive orientation, marked with the red arrows. $r_{\mathrm{in}}$ and $r_{\mathrm{out}}$ have different meanings in different cases. The angles $\Omega_i$ represent the directional change of the vector $\phi^a$ at the joints between adjacent segments $l_i$, used to compute the winding number via Eq. (\ref{W}).}
	\label{C}
\end{figure}

For each circular orbit, one can assign a topological charge. From \cite{Duan:1979ucg,Duan:1984ws,Wei:2020rbh}, the topological current linked to the topological charge is as follows:
\begin{align}
	j^\mu=\frac{1}{2\pi}\epsilon^{\mu\nu\rho}\epsilon_{ab}\partial_\nu n^a\partial_\rho n^b\;,
\end{align}
here, $x^\mu=(t,r,\theta)$ and the unit vector $n^a=(n^r,n^\theta)=(\phi^r/||\phi||,\phi^\theta/||\phi||)$. This current is conserved and can be further simplified to
\begin{align}
	j^\mu=\delta^2(\phi)J^\mu\bigg(\frac{\phi}{x}\bigg)
\end{align}
with the Jacobi tensor given by $\epsilon^{ab}J^\mu\bigg(\frac{\phi}{x}\bigg)=\epsilon^{\mu\nu\rho}\partial_\nu\phi^a\partial_\rho\phi^b$. It is noteworthy that $J^\mu$ only has nonzero values at the zero points of the vector $\phi$.  The  density of the topological current is obtained by denoting the $i$-th zero point as $\vec{x}-\vec{z}_i$.
\begin{align}
	j^0=\sum_{i}^{N}=\beta_i\eta_i\delta^2(\vec{x}-\vec{z}_i)\;,
\end{align}
where the Hopf index ($\beta_i$) and the Brouwer degree ($\eta_i$) of the $i$-th zero point are expressed. The topological number is obtained by integrating the density of the topological current over the given region $\Sigma$
\begin{align}
W=\int_{\Sigma}j^0d^2x=\sum_{i}^{N}\beta_i\eta_i=\sum_{i}^{N}w_i\;,
\end{align}
The symbol $w_i$ signiﬁes the winding number associated with the $i$-th zero point of the vector ﬁeld $\phi$ enclosed within the region $\Sigma$. The topological number can be determined by evaluating the change in direction, denoted as $\Omega$, of the vector as it traverses a counterclockwise closed path $I=\partial\Sigma$
\begin{align}
\label{W}
W=\frac{1}{2\pi}\oint_Id\Omega\;.
\end{align}
Below, following the Ref.~\cite{Wei:2020rbh}, we choose the contour $C=\sum_i\cup l_i$ which encloses $\Sigma$, as shown in Fig.~\ref{C}.


\section{Between two neighboring horizons}\label{section3}
The study of orbital dynamics in the region between multiple horizons, such as the inner Cauchy horizon and the outer event horizon in charged or rotating black holes, presents intriguing theoretical questions. While the practical stability of the inner horizon is debated, the topological approach offers a model-independent way to classify possible orbits in this region.

\begin{figure}[H]
	\centering
	
	\subfigure{
		\begin{minipage}[t]{0.5\linewidth}
			\centering
			\includegraphics[width=2.5in]{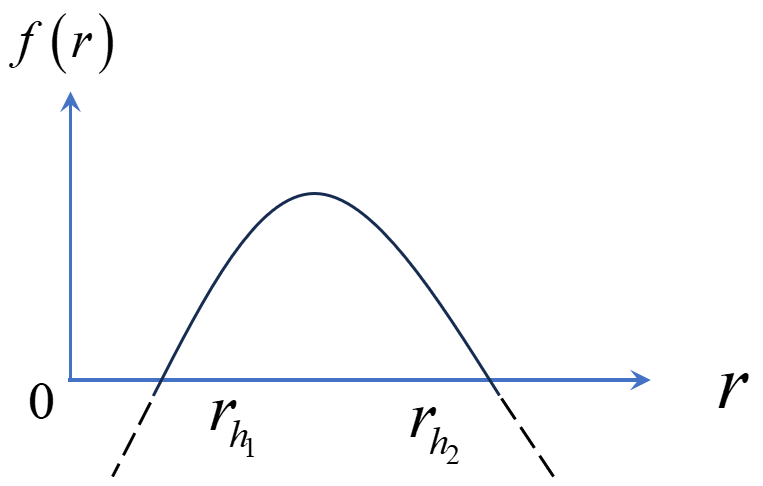}
			\begin{center}
				(a).\, $f'>0$ at $r_{h_1}$, and $f'<0$ at $r_{h_2}$.
			\end{center}
		\end{minipage}%
	}%
	\subfigure{
		\begin{minipage}[t]{0.5\linewidth}
			\centering
			\includegraphics[width=2.5in]{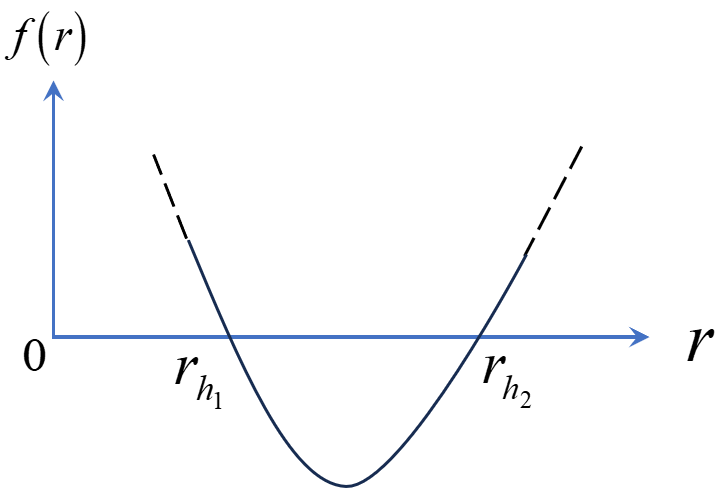}
			\begin{center}
				(b).\, $f'<0$ at $r_{h_1}$, and $f'>0$ at $r_{h_2}$.
			\end{center}		
		\end{minipage}%
	}%
	\centering
	\caption{The behavior of $f(r)$ in the region between two neighboring horizons.}
	\label{fab}
\end{figure}

Assume that there are multiple horizons and consider two  neighboring horizons $r_{h_1}$ and $r_{h_2}$ with $r_{h_1}<r_{h_2}$. The behavior of $f(r)$ is given in Fig.~\ref{fab}. Below, we only consider the situation (a) in Fig.~\ref{fab}. For the situation (b), since $f(r)<0$ between $r_{h_1}$ and $r_{h_2}$ would lead to $\phi^\theta$ and $\phi^r$ being undefined, the circular orbit in this case is forbidden. At the horizon, one has $f(r_h)=g(r_h)=0$, while $\sqrt{g/f}|_{r_h}$ remains finite.


\subsection{massless charged particles}
In this subsection, we study the topology of null circular orbits of massless charged particles between $r_{h_1}$ and $r_{h_2}$. To calculate $W$, one needs to study the boundary behaviors of Eqs.(\ref{phir0}) and (\ref{phit0}).  

Here, we adopt the contour $C$ defined in \cite{Cunha:2020azh,Wei:2020rbh}, where $C=\sum_{i} l_i$ or the union of four line segments $l_1\sim l_4$:$\{r_{\mathrm{out}}=r_{h_2},0\le\theta\le\pi\}\cup\{\theta=\pi, r_{h_1}\le r<r_{h_2}\}\cup\{r_{\mathrm{in}}=r_{h_1},0\le\theta\le\pi\}\cup\{\theta=0,r_{h_1}\le r<r_{h_2}\}$.

Referring to~\cite{Wei:2020rbh}, we study the asymptotic behavior of Eqs. (\ref{phir0}) and (\ref{phit0}). At $r_{h_1}$, we have
\begin{align}
	\phi^r_{l_3}(r\to r_{h_1}^+)>0,\quad \phi^\theta_{l_3}(r\to r_{h_1}^+)\to 0\;,
\end{align}
where $f'(r_{h_1})>0$ is used. The vector $\phi$ is horizontal and points to the left at the horizon $r_{h_1}$, and thus $\Omega_{l_3}=0$.

At $r_{h_2}$, we have
\begin{align}
	\phi^r_{l_1}(r\to r_{h_2}^-)<0,\quad \phi^\theta_{l_1}(r\to r_{h_2}^-)\to 0\;,
\end{align}
where $f'(r_{h_2})<0$ is used. The vector $\phi$ is horizontal  and points to the right at the horizon $r_{h_2}$, and thus $\Omega_{l_1}=\pi$ or $-\pi$.

At $\theta=0$ and $\pi$, we have, respectively,
\begin{align}
	\phi^r_{l_4}(\theta\to 0^+)\sim \frac{1}{\theta},\quad\phi^\theta_{l_4}(\theta\to 0^+)\sim -\frac{1}{\theta^2}\;,
	\phi^r_{l_2}(\theta\to \pi^-)\sim \frac{1}{\pi-\theta},\quad\phi^\theta_{l_2}(\theta\to \pi^-)\sim \frac{1}{(\pi-\theta)^2}\;,
\end{align}
So, we have $\Omega_{l_2}=\frac{\pi}{2}$ and $\Omega_{l_4}=-\frac{\pi}{2}$. 

From the above, we conclude that $\Delta\Omega_1=\Delta\Omega_2=\Delta\Omega_3=\Delta\Omega_4=0$ and $\Omega_1=\Omega_2=\Omega_3=\Omega_4=-\frac{\pi}{2}$. Therefore, we can obtain
\begin{align}
\label{HorizonN}
W=\frac{1}{2\pi}(\Delta\Omega_{l_1}+\Delta\Omega_{l_2}+\Delta\Omega_{l_3}+\Delta\Omega_{l_4}+\Omega_{1}+\Omega_{2}+\Omega_{3}+\Omega_{4})=-1\;.
\end{align}
Based on the results in~\cite{Cunha:2020azh,Wei:2022mzv}, $W=1$ indicates the presence of at least one stable circular orbit, while $W=-1$ indicates that the presence of at least one unstable circular orbit in the region under study. In the above case, for a fixed angular momentum and charge, there is always at least one unstable null circular orbit between $r_{h_1}$ and $r_{h_2}$.


\subsection{massive charged particles}
In this subsection, we investigate the topological properties of circular orbits for massive charged particles located between $r_{h_1}$ and $r_{h_2}$. To do this, we need to examine the boundary behaviors of Eqs. (\ref{phir}) and (\ref{phit}). Following the steps outlined above, one can easily obtain the results for situation (a) in Fig.~\ref{fab}.
\begin{align}
\label{HorizonT}
W=-1\;.
\end{align}
Summary of this section: The topological charge $W=-1$ between two horizons is a direct consequence of the metric function $f(r)$ being positive and having derivatives of opposite signs at the two horizons. This universal result indicates that if circular orbits exist in this region for fixed $(L, q)$, the topology mandates the presence of at least one unstable orbit for both null and timelike cases. This finding extends the topological classification scheme to the inter-horizon region of multi-horizon black holes.

\section{Outside the outermost horizon}\label{section4}
\subsection{Asymptotically flat black holes}
\begin{figure}[H]
	\centering
	\includegraphics[width=3in]{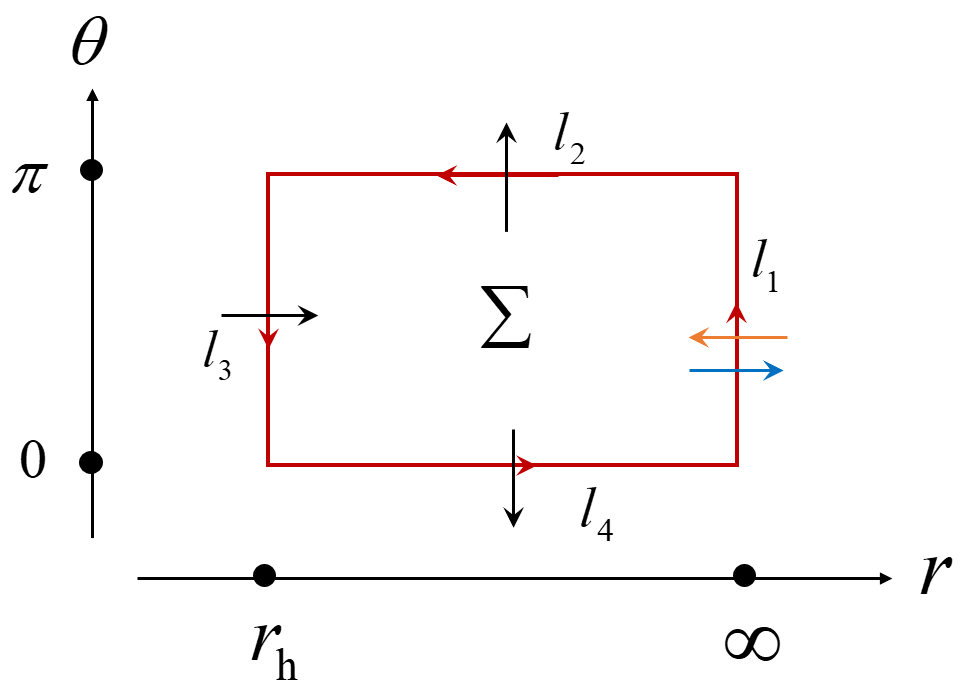}
	\caption{Representation of the contour $C=\sum_il_i$ (which encloses $\Sigma$) on the $(r,\theta)$ plane. The curve $C$ has a positive orientation, marked with the red arrows. The black, bule and yellow arrows indicate the approximate directions of the vector $\phi$ at the boundaries.}
	\label{Cflat}
\end{figure}

In an asymptotically flat black hole, the contour $C=\sum_{i}l_i$ can be defined as: $\{r_{\mathrm{out}}=\infty,0\le\theta\le\pi\}\cup\{\theta=\pi, r_h\le r<\infty\}\cup\{r_{\mathrm{in}}=r_h,0\le\theta\le\pi\}\cup\{\theta=0,r_h\le r<\infty\}$, as shown in Fig.~\ref{Cflat}. It should be noted that here $r_h$ represents the outermost horizon.

\begin{figure}[H]
	\centering
	\includegraphics[width=3in]{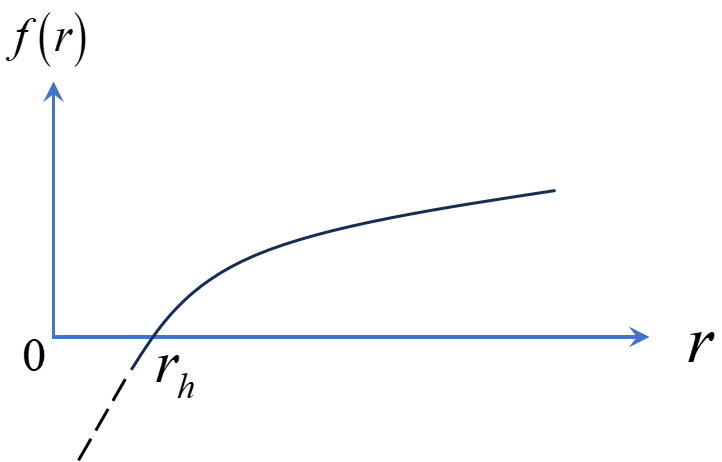}
	\caption{The behavior of $f(r)$  in an asymptotically flat black hole. At $r_h$, it is observed that $f'(r)>0$.}
	\label{fflat}
\end{figure}

In an asymptotically ﬂat black hole with the solution described by (\ref{metric}), the metric functions exhibit the following asymptotic behaviors at $r\to \infty$~\cite{Wei:2020rbh}:
\begin{align}
	\label{Aflat}
	f\sim 1-\frac{2M}{r}+\mathcal{O}\bigg(\frac{1}{r^2}\bigg)\;,\quad g\sim 1-\frac{2M}{r}+\mathcal{O}\bigg(\frac{1}{r^2}\bigg)\;,\quad h\sim r^2\;.
\end{align}
where $M$ is the black hole mass, which is always positive. For $r_h<r<\infty$, $f(r)$ behaves as shown in Fig.~\ref{fflat}, and $f'(r)>0$ at $r_h$.


\subsubsection{massless charged particles}
At the horizon, from Eqs.(\ref{phir0}) and (\ref{phit0}), we obtain
\begin{align}
	\phi^r_{l_3}(r\to r_h)>0,\quad \phi^\theta_{l_3}(r\to r_h)\to 0\;,
\end{align}
where we have used the fact that $f'(r_h)>0$. Consequently, the vector $\phi$  is oriented horizontally to the right at the horizon, which implies $\Omega_{l_3}=0$. At $\theta=0$ and $\pi$, the results are consistent with those in the previous section.

Lastly, let us consider $\Omega$ along $l_1$. When $r\to\infty$,
\begin{align}
\phi^r_{l_1}(r\to\infty)\sim -\frac{q Q+L}{r^2}
\end{align}
on the equatorial plane.

If $-q Q<L$, then $\phi^r_{l_1}(r\to\infty)<0$. As a result, by ignoring the specific value of $\phi^\theta$, the direction of the vector $\phi$ is to the left. As shown in Fig.~\ref{Cflat}, the direction of the vector is outward at $\theta=0$ and $\pi$, and to the right at $r=r_h$, and to the left at $\infty$ (denoted by a yellow arrow), ignoring any possible inclination. According to the asymptotic behaviors of $\phi$, one can easily obtain
\begin{align}
\label{flatNQ_1}
W=\frac{1}{2\pi}\oint d\Omega=\frac{1}{2\pi}\times(-2\pi)=-1\;,
\end{align}
which means that for a fixed angular momentum and charge, there will always be at least one unstable null circular orbit outside the event horizon in this case.

If $-q Q> L$, then $\phi^r_{l_1}(r\to\infty)>0$, and the direction of the vector $\phi$ is to the right at $\infty$ (denoted by a blue arrow in Fig.~\ref{Cflat}), ignoring any possible inclination. According to the asymptotic behaviors of $\phi$, one can easily get
\begin{align}
\label{flatNQ_0}
W=\frac{1}{2\pi}\oint d\Omega=\frac{1}{2\pi}\times(-\pi)+\frac{1}{2\pi}\times(\pi)=0\;,
\end{align}
which means that for a fixed angular momentum and charge, if null circular orbits exist in this case, they must occur in pairs, with one being stable and the other being unstable.

\subsubsection{massive charged paticles}
At the horizon, it is easy to obtain
\begin{align}
	\phi^r_{l_3}(r\to r_h)>0,\quad \phi^\theta_{l_3}(r\to r_h)\to 0\;,
\end{align}
where $f'(r_h)>0$ is used. Consequently, the vector $\phi$ is oriented horizontally to the right at the horizon, which implies $\Omega_{l_3}=0$. 

At $\theta=0$ and $\pi$, we respectively have
\begin{align}
	\phi^r_{l_4}(\theta\to 0^+)\sim \frac{1}{\theta},\quad\phi^\theta_{l_4}(\theta\to 0^+)\sim -\frac{1}{\theta^2}\;,
	\phi^r_{l_2}(\theta\to \pi^-)\sim \frac{1}{\pi-\theta},\quad\phi^\theta_{l_2}(\theta\to \pi^-)\sim \frac{1}{(\pi-\theta)^2}\;,
\end{align}
Thus, we have $\Omega_{l_2}=\frac{\pi}{2}$ and $\Omega_{l_4}=-\frac{\pi}{2}$. 

Lastly, let us consider $\Omega$ along $l_1$. At $r\to\infty$, one have
\begin{align}
\phi^r_{l_1}(r\to\infty)\sim \frac{M-q Q}{r^2}
\end{align}
on the equatorial plane. 

If $M>q Q$, $\phi^r_{l_1}(r\to\infty)>0$, and the direction of the vector $\phi$ is to the right at $\infty$ (denoted by a blue arrow in Fig.~\ref{Cflat}), ignoring any possible inclination. According to the asymptotic behaviors of $\phi$, one can easily get
\begin{align}
	\label{flatTQ_0}
W=\frac{1}{2\pi}\oint d\Omega=\frac{1}{2\pi}\times(-\pi)+\frac{1}{2\pi}\times(\pi)=0\;,
\end{align}
which means that for a fixed angular momentum and charge, if TCOs exist outside the outermost horizon in this case, they must occur in pairs, with one being stable and the other being unstable.

If $M< q Q$, $\phi^r_{l_1}(r\to\infty)<0$, and the direction of the vector $\phi$ is to the left at $\infty$ (denoted by a yellow arrow in Fig.~\ref{Cflat}), ignoring any possible inclination. According to the asymptotic behaviors of $\phi$, one can easily get
\begin{align}
	\label{flatTQ_1}
W=\frac{1}{2\pi}\oint d\Omega=\frac{1}{2\pi}\times(-\pi)+\frac{1}{2\pi}\times(-\pi)=-1\;,
\end{align}
which means that for a fixed angular momentum and charge, there will always exist at least one unstable TCO located outside the event horizon in this case. These results are consistent with those in Ref.~\cite{Ye:2024sus}.


\subsubsection{RN black hole}
\begin{figure}[H]
	\centering
	\includegraphics[width=3.5in]{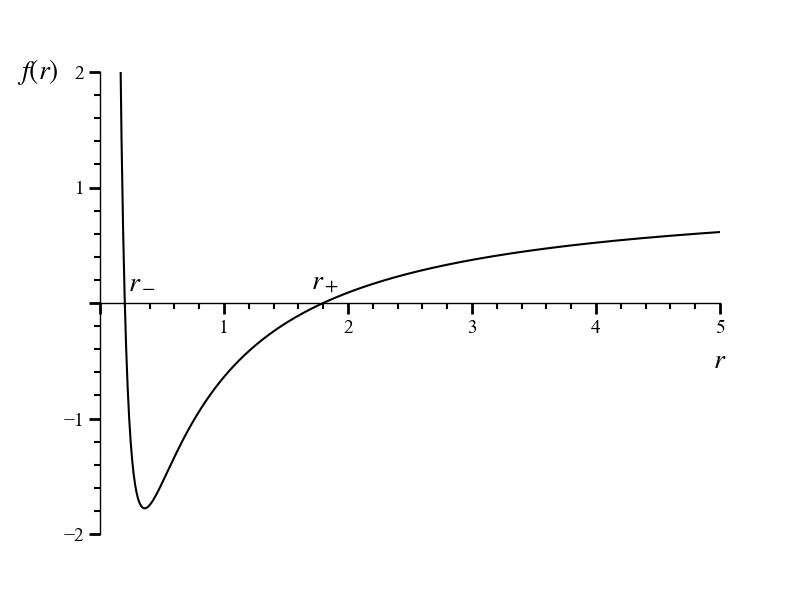}
	\caption{The behavior of $f(r)$ in RN black hole. At $r_-$, one has $f'(r)<0$; at $r_+$, one has $f'(r)>0$.}
	\label{fRN}
\end{figure}

We use Reissner-Nordstr$\ddot{\mathrm{o}}$m (RN) black holes as an example to verify the previous conclusions. Consider a non-extreme RN black hole, where the functions $f(r)$, $g(r)$ and $h(r)$ in (\ref{metric}) are
\begin{align}
	f(r)=g(r)=1-\frac{2M}{r}+\frac{Q^2}{r^2}\;,\quad h(r)=r^2\;.
\end{align}
Here, $M>Q$. There are two horions $r_-$ and $r_+$, defined by $f(r_-)=f(r_+)=0$, and
\begin{align}
	r_{\pm}=M\pm\sqrt{M^2-Q^2}\;.
\end{align}
The associated electomagnetic potential is
\begin{align}
	A=-\frac{Q}{r}dt\;.
\end{align}
Below, we set $M=1$ and $Q=0.6$. The graph of the function $f(r)$ is shown in Fig.~\ref{fRN}, with $r_-=0.2$, $r_+=1.8$.
For the region $r_-$ to $r_+$, there are no circular orbits. Therefore, we focus on the case where $r>r_+$. By analyzing the graph of the potential function, we can determine whether circular orbits exist and assess their stability. In the following analysis, we will concentrate on the graphical representation of the potential function. The number of extrema on the graph corresponds to the number of circular orbits. Specifically, a local maximum on the graph signifies an unstable circular orbit, while a local minimum indicates a stable circular orbit. Below, we will use 3D graphs to observe the changes in the extremum points of the potential function\footnote{In the creation of 3D graphs, visual offsets in the labeling lines occur because they are drawn at the minimum Z-height rather than following the actual surface contour. This issue is compounded by perspective projection effects. To avoid misleading readers, we have employed multiple perspectives and included specific annotations in the figure.}.

\begin{figure}[H]
	\centering
	\includegraphics[width=7in]{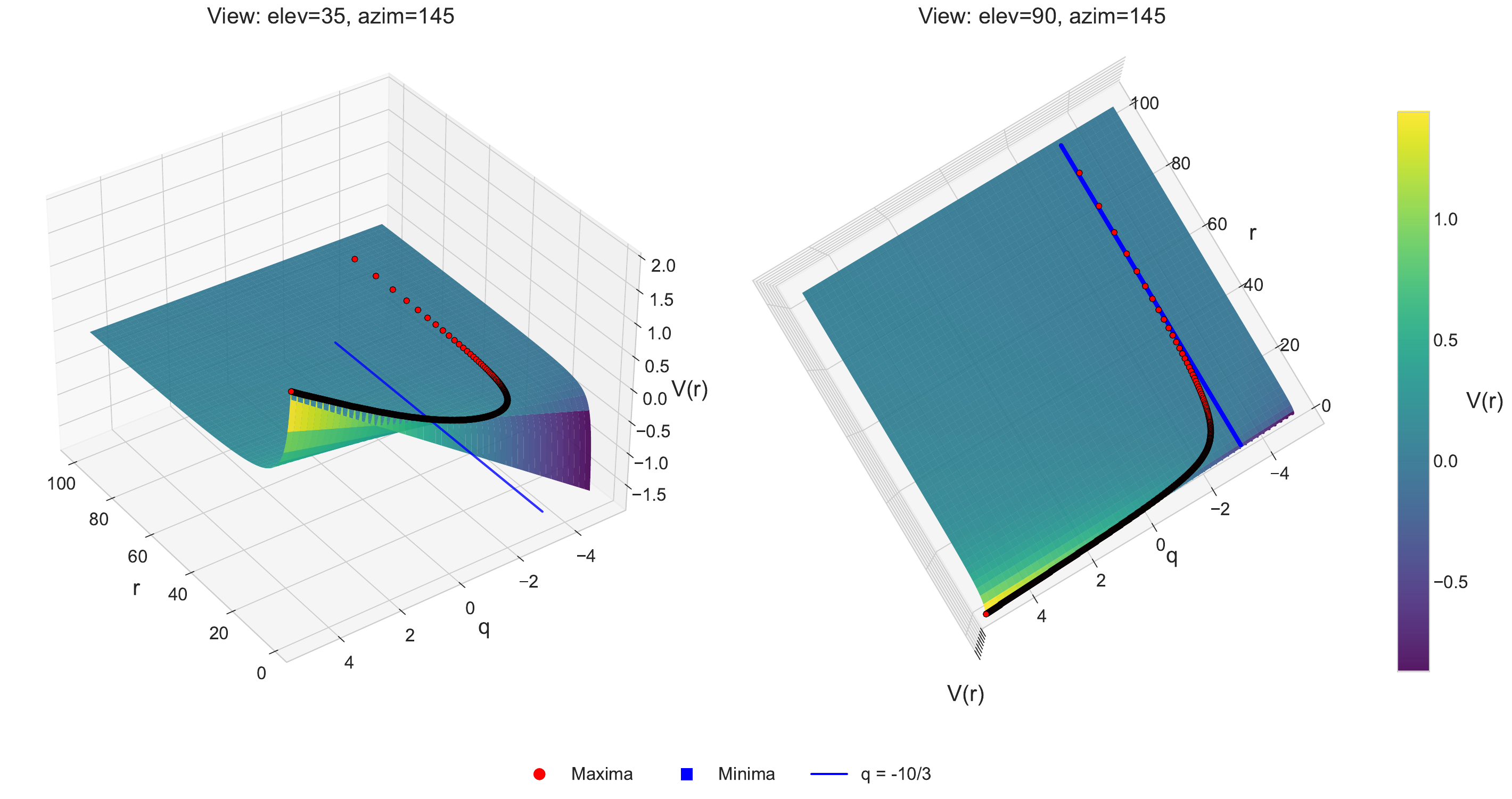}
	\caption{The graph of the potential function (\ref{VflatN}) for $-5\le q\le 5$ and $L=2$. The red points represent the maximum values of (\ref{VflatN}) for a given value of $q$. The blue line, which corresponds to $q=-10/3$, indicates where the extremum points disappear.}
	\label{RNN}
\end{figure}

Due to the spherical symmetry of the system, we can study the circular motion of a charged particle confined to the equatorial plane. From equation (\ref{V}), we derive the effective potential of the particle on the equatorial plane as follows:
\begin{align}
	V=V_+=\frac{q Q}{r}+\sqrt{\bigg(1-\frac{2M}{r}+\frac{Q^2}{r^2}\bigg)\bigg(\mu^2+\frac{L^2}{r^2}\bigg)}\;.
\end{align}
For a massless charged particle, we have
\begin{align}
	\label{VflatN}
	V=\frac{q Q}{r}+\sqrt{\bigg(1-\frac{2M}{r}+\frac{Q^2}{r^2}\bigg)\frac{L^2}{r^2}}\;,
\end{align}
Consider $-5\le q\le 5$ and $L=2$. The graph of the potential function (\ref{VflatN}) is shown in Fig~\ref{RNN}. From this figure, when $q>L/Q=-10/3$, there exists an unstable null circular orbit, which is consistent with the result (\ref{flatNQ_1}). When $q<L/Q=-10/3$, there are no null circular orbits, which is consistent with the result (\ref{flatNQ_0}).

\begin{figure}[H]
	\centering
	\includegraphics[width=7in]{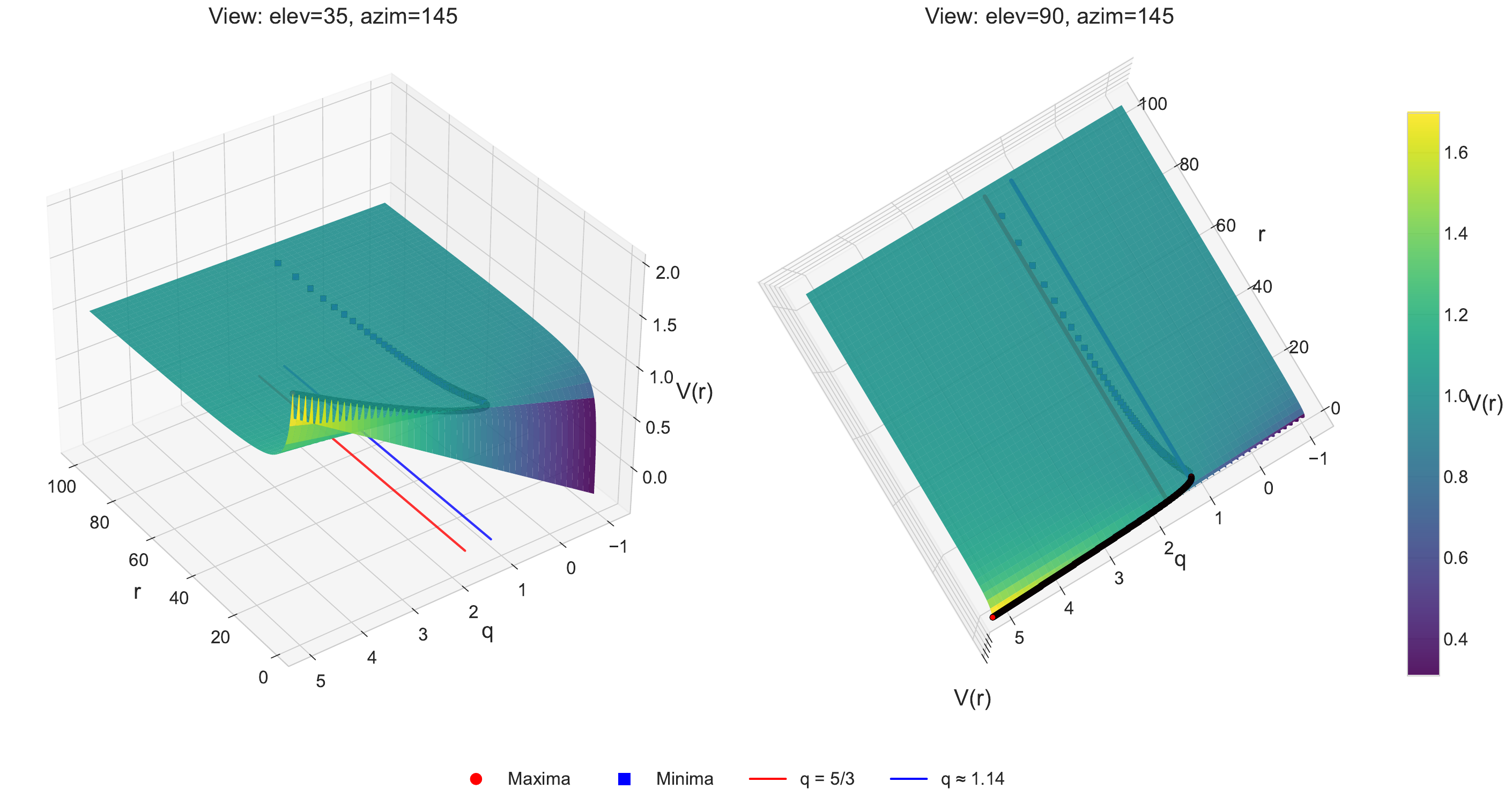}
	\caption{The graph of the potential function (\ref{VflatT}) for $-1\le q\le 5$ and $L=2$. The red points and the blue boxes represent the maximum values and the minimum values of (\ref{VflatT}) for a given value of $q$, respectively. The red line, which corresponds to $q=5/3$, represents the point where the number of extremum points changes. The blue line, which corresponds to $q\approx1.14$, indicates where the extremum points disappear.}
	\label{RNT}
\end{figure}

For a massive charged particle, we have
\begin{align}
\label{VflatT}
	V=\frac{q Q}{r}+\sqrt{\bigg(1-\frac{2M}{r}+\frac{Q^2}{r^2}\bigg)\bigg(1+\frac{L^2}{r^2}}\bigg)\;,
\end{align}
Consider $-1\le q\le 5$ and $L=2$. The graph of the potential function (\ref{VflatT}) is shown in Fig.~\ref{RNT}. From this figure, when $q>M/Q=5/3$, there exists an unstable TCO, which is consistent with the result (\ref{flatTQ_1}). When $1.14<q<M/Q=5/3$, there is one stable TCO and one unstable TCO. When $q<1.14$, there are no TCOs. These results are all consistent with the result (\ref{flatTQ_0})\footnote{From Eq.(\ref{VflatT}) and by requiring $V'(r)=V''(r)=0$, one can obtain $q\approx1.14$, where the extremum points disappear.}.


\subsection{Asymptotically AdS black holes}
\begin{figure}[H]
	\centering
	\includegraphics[width=2.5in]{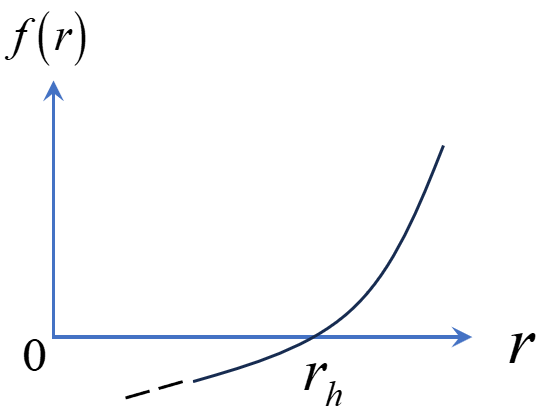}
	\caption{The behavior of $f(r)$ in asymptotically AdS black hole, and $f'(r)>0$ at $r_h$.}
	\label{fAdS}
\end{figure}

\begin{figure}[H]
	\centering
	\includegraphics[width=3in]{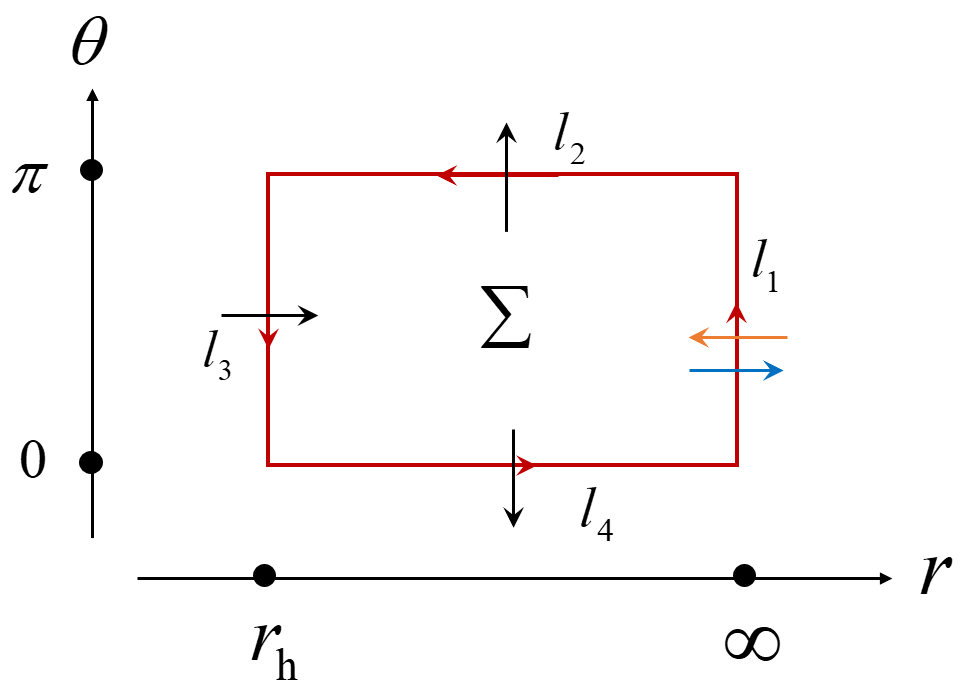}
	\caption{The representation of the contour $C=\sum_i\cup l_i$ (which encloses $\Sigma$) on the $(r,\theta)$ plane. The curve $C$ has a positive orientation marked with the red arrows. The black, blue and yellow arrows indicate the approximate direction of the vector $\phi$ along the boundaries.}
	\label{CAdS}
\end{figure}

An asymptotically AdS black hole exhibits the following asymptotic behavior~\cite{Wei:2020rbh}:
\begin{align}
	\label{AAdS}
	&f(r)\sim \frac{r^2}{l^2}+1-\frac{2M}{r}+\mathcal{O}\bigg(\frac{1}{r^2}\bigg)\;,\nonumber\\
	&g(r)\sim \frac{r^2}{l^2}+1-\frac{2M}{r}+\mathcal{O}\bigg(\frac{1}{r^2}\bigg)\;,\nonumber\\
	&h(r)\sim r^2\;,
\end{align}
where $l$ is the AdS radius. For $r_h<r<\infty$,  the function $f(r)$ exhibits the behavior shown in Fig.~\ref{fAdS}, and $f'(r)>0$ at $r_h$. Note that these asymtotical behaviors only modify the boundary condition at $r\to\infty$, so the calculation along $l_2$, $l_3$ and $l_4$ is the same as the asymptotically flat case.
The contour we choose in asymptotically AdS spacetimes is shown in Fig.~\ref{CAdS}, which is the same as that in asymptotically flat spacetimes. Therefore, the same boundary behavior will yield the same results.

\subsubsection{massless charged particles}
When $r\to\infty$, we have a finite $l$, and
\begin{align}
\phi_{l_1}^r(r\to\infty)=-\frac{q Q}{rl}
\end{align}
on the equatorial plane.

If $q Q> 0$, one has $\phi_{l_1}^r(r\to\infty)<0$. Following the results in asymtotically flat spacetimes, the total topological number is
\begin{align}
\label{RNAdSNQ_1}
W=-1\;,
\end{align}
which indicates that for a fixed angular momentum and charge, there will always be at least one unstable null circular orbit outside the event horizon in this case.

If $q Q<0$, one have $\phi_{l_1}^r(r\to\infty)>0$. Following the results in asymtotically flat spacetimes, the total topological number is
\begin{align}
	\label{RNAdSNQ0}
W=0\;,
\end{align}
which means that for a fixed angular momentum and charge, if null circular orbits exist in this case, they must occur in pairs, with one being stable and the other unstable.

\subsubsection{massive charged particles}
When $r\to\infty$, we have a finite $l$, and 
\begin{align}
\phi_{l_1}^r(r\to\infty)=\frac{r}{l^2}\;,
\end{align}
Thus, we will always have $\phi_{l_1}^r(r\to\infty)>0$. Following the results in asymptotically flat spacetimes, the total topological number is
\begin{align}
	\label{RNAdSTQ0}
	W=0\;,
\end{align}
which means that if TCOs with fixed angular momentum and charge exist in asymptotically AdS spacetime, they must occur in pairs, with one being stable and the other unstable.

\subsubsection{RN-AdS black hole}
\begin{figure}[H]
	\centering
	\includegraphics[width=3.5in]{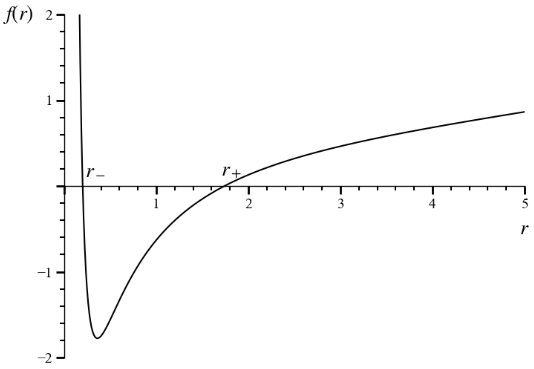}
	\caption{The behavior of $f(r)$ in RN-AdS black hole. At $r_-$, one has $f'(r)<0$; at $r_+$, one has $f'(r)>0$.}
	\label{fRNAdS}
\end{figure}

\begin{figure}[H]
	\centering
	\includegraphics[width=7in]{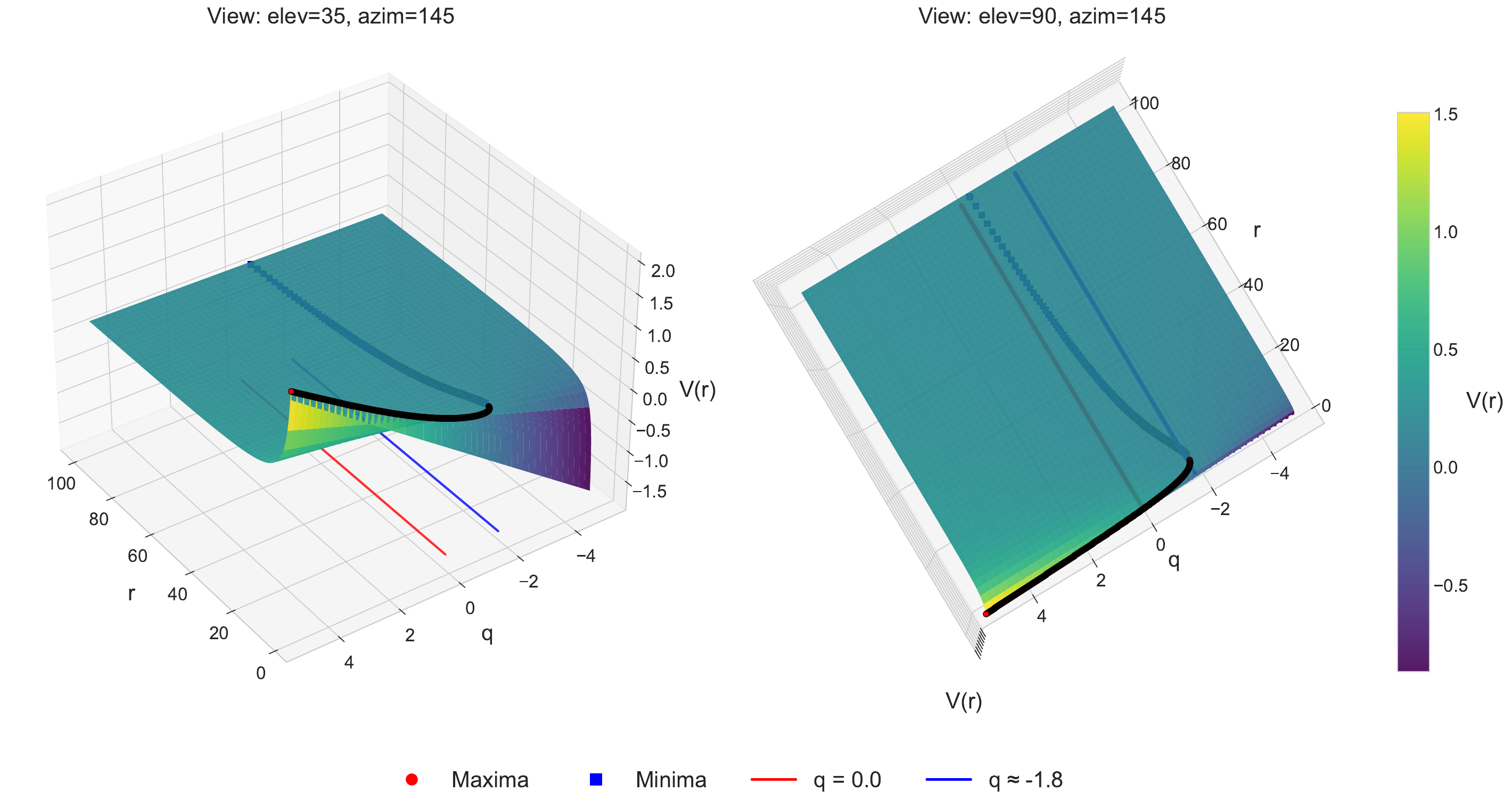}
	\caption{The graph of the potential function (\ref{VAdSN}) for $-5\le q\le 5$ and $L=2$. The red points and the blue boxes represent the maximum values and the minimum values of (\ref{VAdSN}) for a given value of $q$, respectively. The red line, which corresponds to $q=0$, represents the point where the number of extremum points changes. The blue line, which corresponds to $q\approx-1.8$, indicates where the extremum points disappear.}
	\label{RNAdSN}
\end{figure}

\begin{figure}[H]
	\centering
	\includegraphics[width=7in]{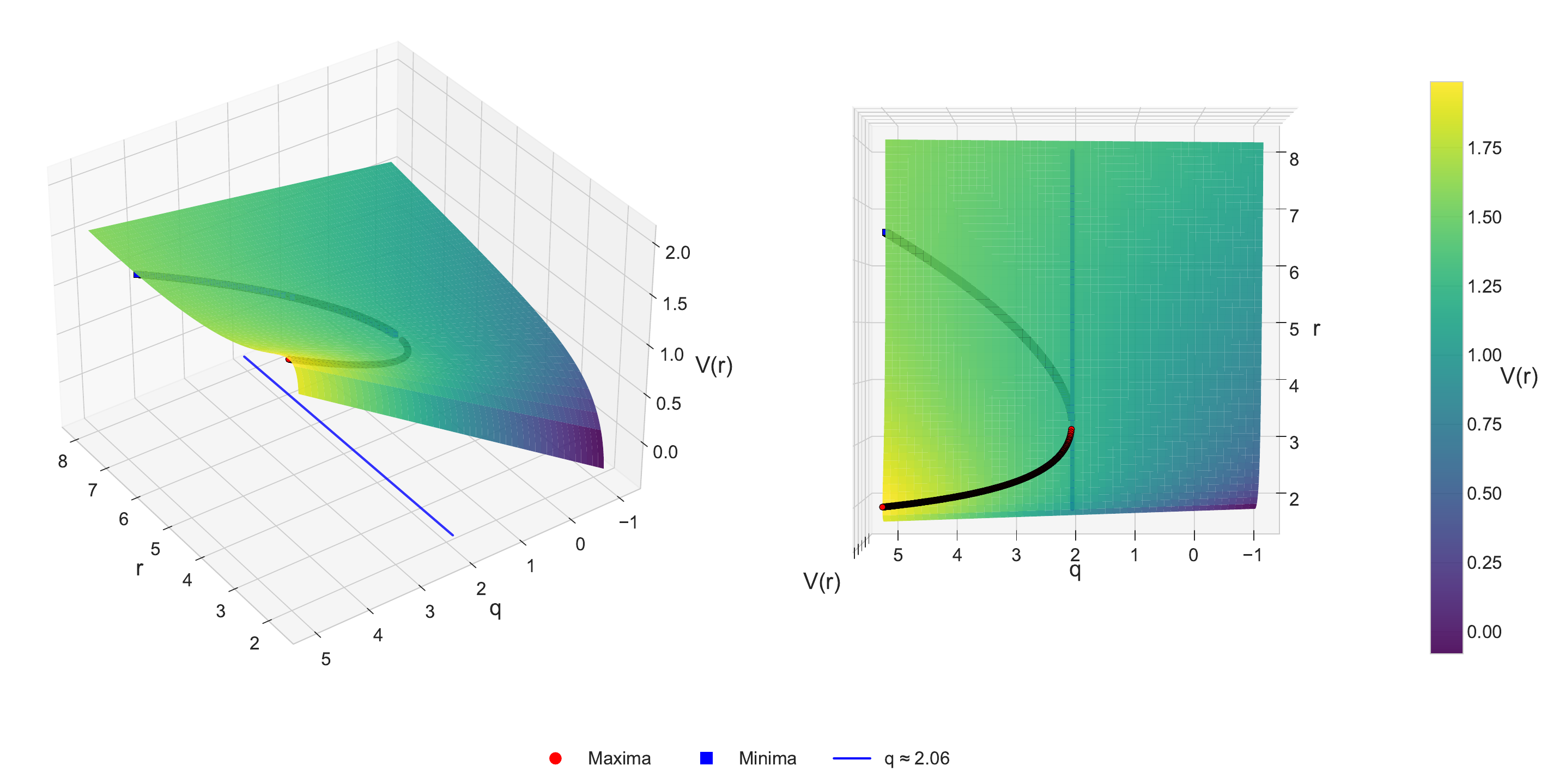}
	\caption{The graph of the potential function (\ref{VAdSN}) for $-1\le q\le 5$ and $L=2$. The red points and the blue boxes represent the maximum values and the minimum values of (\ref{VAdSN}) for a given value of $q$, respectively. The blue line, which corresponds to $q\approx2.06$, indicates where the extremum points disappear.}
	\label{RNAdST}
\end{figure}

Here, we use Reissner-Nordstr$\ddot{\mathrm{o}}$m Anti-de Sitter (RN-AdS) black holes as an example to verify the above conclusions. For a RN-AdS black hole, the functions $f(r)$, $g(r)$ and $h(r)$ in (\ref{metric}) are given by~\cite{Ma:2016vop}
\begin{align}
	f(r)=g(r)=1-\frac{2M}{r}+\frac{Q^2}{r^2}+\frac{r^2}{l^2}\;,\quad h(r)=r^2\;.
\end{align}
where $l$ is the AdS length scale. The associated electomagnetic potential is
\begin{align}
	A=-\frac{Q}{r}dt\;.
\end{align}
Below, we set $M=1$, $Q=0.6$ and $l=10$. The graph of the function $f(r)$ is shown in Fig.~\ref{fRNAdS}, with $r_-=0.20001$, $r_+=1.74044$. In the region between $r_-$ and $r_+$, there are no circular orbits. Therefore, we focus on the case where $r>r_+$.
For a massless charged particle, we have
\begin{align}
	\label{VAdSN}
	V=\frac{q Q}{r}+\sqrt{\bigg(1-\frac{2M}{r}+\frac{Q^2}{r^2}+\frac{r^2}{l^2}\bigg)\frac{L^2}{r^2}}\;,
\end{align}
Consider $-5\le q\le 5$ and $L=2$. The graph of the potential function (\ref{VAdSN}) is shown in Fig.~\ref{RNAdSN}. From this figure, when $q>0$, there exists an unstable null circular orbit, which is consistent with the result (\ref{RNAdSNQ_1}). When $-1.8<q<0$, there is one stable null circular orbit and one unstable null circular orbit. When $q<-1.8$, there are no null circular orbits. These results are all consistent with the result (\ref{RNAdSNQ0}).
For a massive charged particle, we have
\begin{align}
	\label{VAdST}
	V=\frac{q Q}{r}+\sqrt{\bigg(1-\frac{2M}{r}+\frac{Q^2}{r^2}+\frac{r^2}{l^2}\bigg)\bigg(1+\frac{L^2}{r^2}}\bigg)\;,
\end{align}
Consider $-1\le q\le 5$ and $L=2$. The graph of the potential function (\ref{VAdST}) is shown in Fig.~\ref{RNAdST}. From this figure, when $q>2.06$, there is one stable TCO and one unstable TCO. When $q<2.06$, there are no TCOs. These results are consistent with (\ref{RNAdSTQ0}).


\subsection{Asymptotically dS black holes}
\begin{figure}[H]
	\centering
	\includegraphics[width=3in]{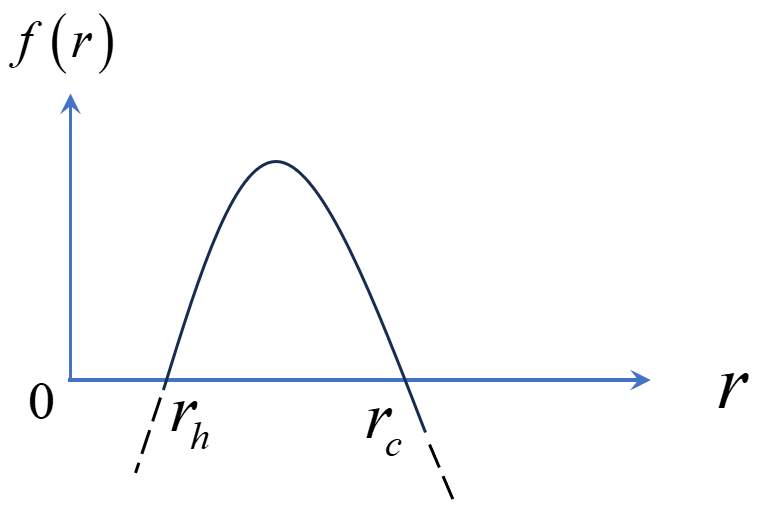}
	\caption{The behavior of $f(r)$ in asymptotically dS black hole. One has $f'(r)>0$ at $r_h$ and $f'(r)<0$ at $r_c$. Here, $r_h$ denotes the outermost black hole horizon, while $r_c$ represents the cosmological horizon.}
	\label{fdS}
\end{figure}

Now we turn to the asymtotically dS black hole case, which exhibits the following asymptotic behavior at large $r$~\cite{Wei:2020rbh}
\begin{align}
	&f(r)\sim -\frac{r^2}{l^2}+1-\frac{2M}{r}+\mathcal{O}\bigg(\frac{1}{r^2}\bigg)\;,\\
	&g(r)\sim -\frac{r^2}{l^2}+1-\frac{2M}{r}+\mathcal{O}\bigg(\frac{1}{r^2}\bigg)\;,\\
	&h(r)\sim r^2\;,
\end{align}
where $l$ is the curvature radius of dS spacetime. Unlike the asymptotically flat and AdS cases, besides the black hole horizon, the spacetime admits a cosmological horizon with radius $r_c>r_h$, which is also a root of $g(r_c)=f(r_c)=0$. Therefore, for this case, we restrict our consideration to $r_h\le r\le r_c$. The behavior of $f(r)$ within $r_h\le r\le r_c$ is shown in Fig.~\ref{fdS}, which corresponds exactly to situation (a) in Fig.~\ref{fab}.
Thus, one can easily derive
\begin{align}
\label{RNdSQ_1}
W=-1
\end{align}
for both massless charged particles and massive charged particles. This implies that between $r_h$ and $r_c$ in an asymptotically dS black hole, for a fixed angular momentum and charge, there will always be at least one unstable null circular orbit and one unstable TCO.


\subsubsection{RN-dS black hole}
\begin{figure}[H]
	\centering
	\includegraphics[width=3.5in]{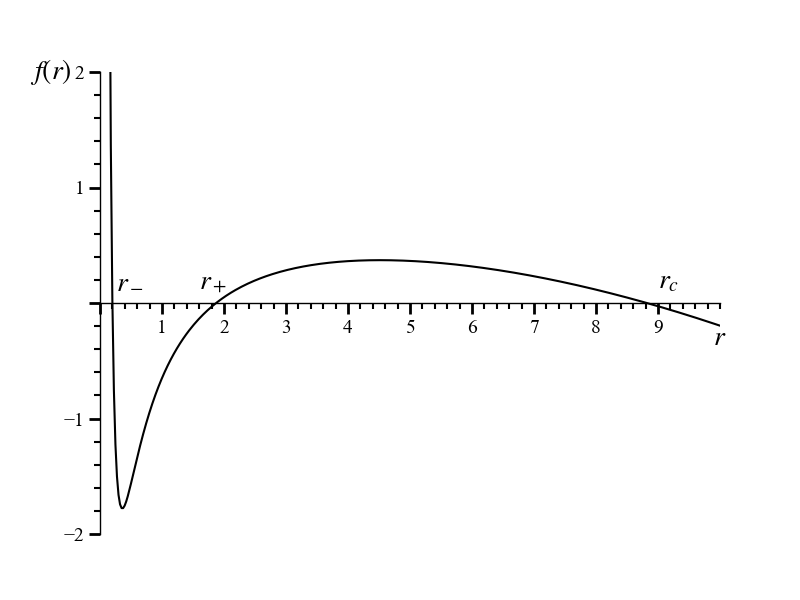}
	\caption{The behavior of $f(r)$ in the RN-dS black hole is such that $f'(r)>0$ at $r_+$, while $f'(r)<0$ at both $r_-$ and $r_c$.}
	\label{fRNdS}
\end{figure}

\begin{figure}[H]
	\centering
	\includegraphics[width=4in]{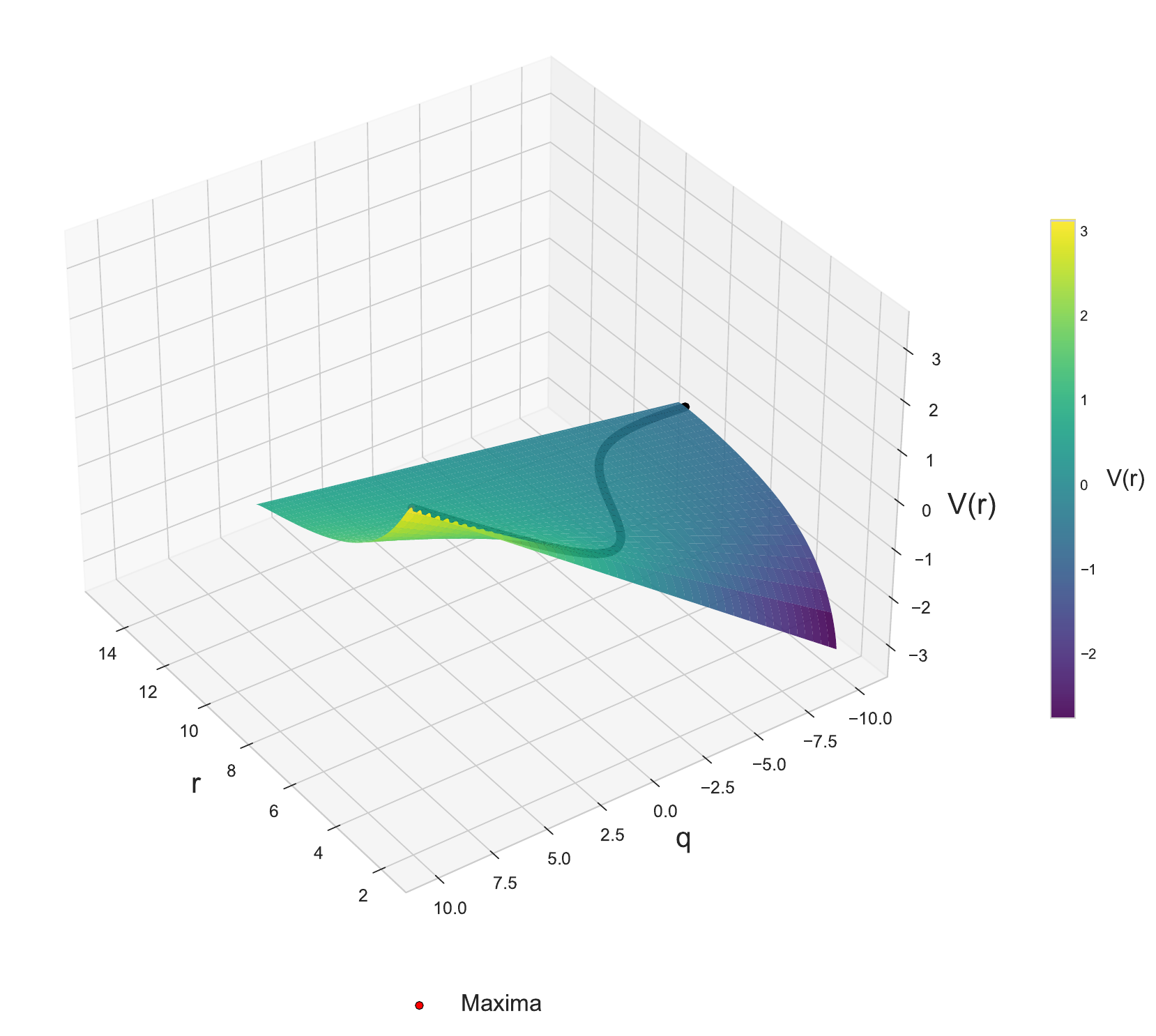}
	\caption{The graph of the potential function (\ref{VdSN}) for $-10\le q\le 10$ and $L=2$. The red points represent the maximum values of (\ref{VdSN}) for a given value of $q$.}
	\label{RNdSN}
\end{figure}

\begin{figure}[H]
	\centering
	\includegraphics[width=4in]{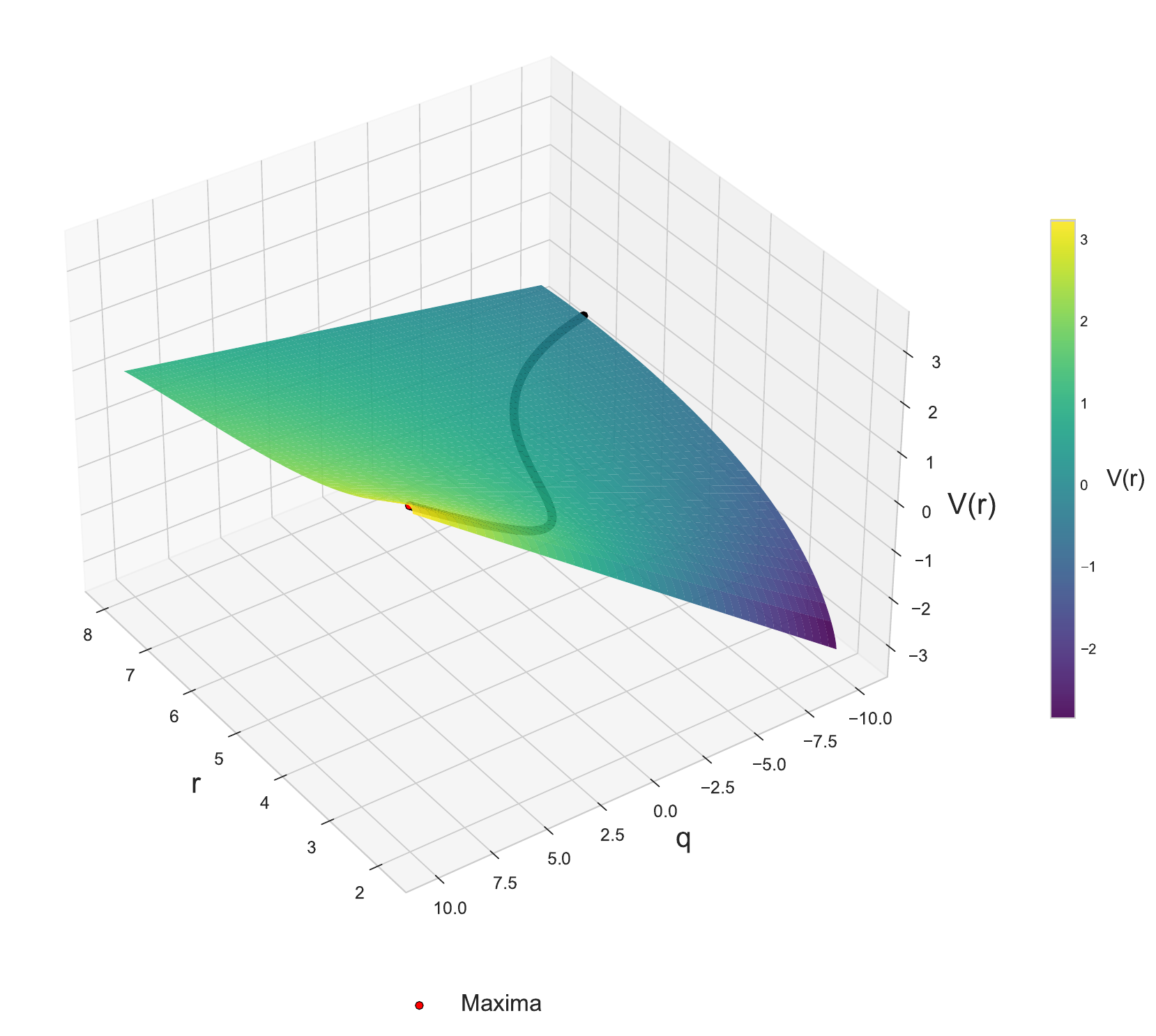}
	\caption{The graph of the potential function (\ref{VdST}) for $-10\le q\le 10$ and $L=2$. The red points represent the maximum values of (\ref{VdST}) for a given value of $q$.}
	\label{RNdST}
\end{figure}

Here, we use Reissner-Nordstr$\ddot{\mathrm{o}}$m de Sitter black holes (RN-dS) as an example to verify the above conclusions. For a RN-dS black hole, the functions $f(r)$, $g(r)$ and $h(r)$ in (\ref{metric}) are given by~\cite{Du:2022jcb}
\begin{align}
	f(r)=g(r)=1-\frac{2M}{r}+\frac{Q^2}{r^2}-\frac{r^2}{l^2}\;,\quad h(r)=r^2\;.
\end{align}
The associated electomagnetic potential is
\begin{align}
	A=-\frac{Q}{r}dt\;.
\end{align}
Below, we set $M=1$, $Q=0.6$ and $l=10$. The graph of the function $f(r)$ is shown in Fig.~\ref{fRNdS}, with $r_-=0.3674$, $r_+=1.6946$ and $r_c=8.83985$. In the region between $r_-$ and $r_+$, there are no circular orbits. Therefore, we focus on the case where $r_+<r<r_c$.
For a massless charged particle, we have
\begin{align}
	\label{VdSN}
	V=\frac{q Q}{r}+\sqrt{\bigg(1-\frac{2M}{r}+\frac{Q^2}{r^2}-\frac{r^2}{l^2}\bigg)\frac{L^2}{r^2}}\;,
\end{align}
Consider $-10\le q\le 10$ and $L=2$. The graph of the potential function (\ref{VdSN}) is shown in Fig.~\ref{RNdSN}. From this figure, it is evident that there is always an unstable null circular orbit, which is consistent with the result (\ref{RNdSQ_1}).
For a massive charged particle, we have
\begin{align}
	\label{VdST}
	V=\frac{q Q}{r}+\sqrt{\bigg(1-\frac{2M}{r}+\frac{Q^2}{r^2}-\frac{r^2}{l^2}\bigg)\bigg(1+\frac{L^2}{r^2}}\bigg)\;,
\end{align}
Consider $-10\le q\le 10$ and $L=2$. The graph of the potential function (\ref{VdST}) is shown in Fig.~\ref{RNdST}. From this figure, it is evident that there is always an unstable TCO, which is consistent with the result (\ref{RNdSQ_1}).


\section{Conclusions and discussions}\label{section5}
Although topological studies of circular orbits in spacetime cannot yield the exact number of such orbits, they can nevertheless reveal global properties of these orbits and link the winding number to orbital stability (where $W=1$ corresponds to stable circular orbits, while $W=-1$ corresponds to unstable circular orbits). This global perspective not only deepens our understanding of the orbits themselves but also supplies new analytical tools for other branches of physics~\cite{Wei:2021vdx,Wei:2022dzw}.

In this study, we have systematically investigated the topological properties of circular orbits for charged test particles in asymptotically flat, anti-de Sitter (AdS), and de Sitter (dS) black holes with multiple horizons. By employing a topological approach based on vector field analysis and winding number calculations, we have uncovered the influence of particle charge and spacetime asymptotic behavior on the existence and stability of both timelike and null circular orbits. Our findings are further validated through explicit examples of Reissner-Nordström (RN), RN-AdS, and RN-dS black holes. The key conclusions are summarized in Tab.~\ref{Summary}:
\begin{table}[htb]
	\centering
	\caption{For V=$V_+$, the topology of circular orbits for charged particles in spacetime.}
	\includegraphics[width=5in]{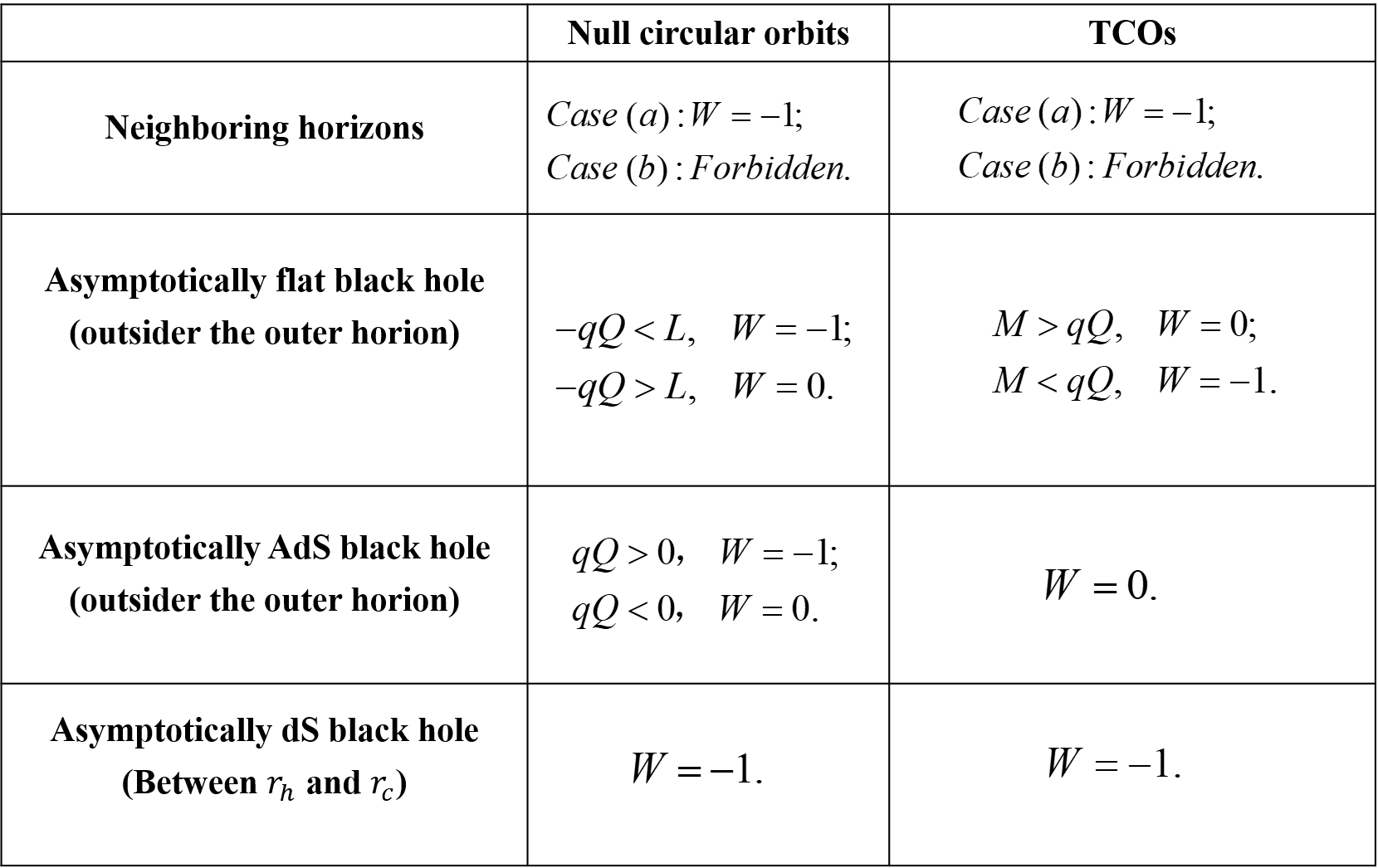}
	\label{Summary}
\end{table}
\begin{itemize}
	\item Multi-horizon black holes: For black holes with multiple horizons, if circular orbits with fixed angular momentum and charge exist between two neighboring horizons (Case (a) in Fig.~\ref{fab}), there must always exist at least one unstable null circular orbit and one unstable TCO. Conversely, regions where $f(r)<0$ (Case (b)) entirely forbid such circular orbits.
\end{itemize} 

\begin{itemize}
	\item Asymptotically flat black holes: For null circular orbits with fixed angular momentum and charge, the topological number $W$ depends on the charge $(
	q Q)$ and angular momentum $(L)$ of the charged particle. When $-q Q<L$, $W=-1$, indicating there exists at least one unstable null circular orbit. If $-q Q> L$, $W=0$, implying paired null circular orbits (one stable, one unstable) or no null circular orbits at all.
	
	For TCOs with fixed angular momentum and charge, the black hole mass ($M$) and the charge ($q Q$) determine $W$. When $M>q Q$, $W=0$ implying paired TCOs (one stable, one unstable) or no circular orbits at all. If $M < q Q$, $W=-1$, indicating there exists at least one unstable TCO.
\end{itemize}

\begin{itemize}
	\item Asymptotically AdS black holes: Null circular orbits with fixed angular momentum and charge exhibit $W=-1$ for $q Q> 0$, guaranteeing the existence of an unstable null circular orbit. For $q Q<0$, $W=0$, permitting either paired null circular orbits (one stable and one unstable) or no null circular orbits at all. TCOs with fixed angular momentum and charged, universally yield $W=0$, requiring paired TCOs when they exist, independent of the particle's charge.
\end{itemize}

\begin{itemize}
	\item Asymptotically dS black holes: Between the outermost event horizon ($r_h$) and the cosmological horizon ($r_c$), irrespective of particle's charge, there must always exist at least one unstable null circular orbit and one unstable TCO for a fixed angular momentum and charge.
\end{itemize}
These results highlight the critical role of spacetime asymptotics and particle charge in shaping the topological landscape of circular orbits. Notably, the charge of a particle introduces additional constraints on orbit stability, distinguishing this work from previous studies focused on neutral particles.

While the net charge of astrophysical black holes is expected to be negligible, the dynamics studied here could be relevant in effective descriptions. For instance, in magnetized accretion plasmas, particles may experience effective charge-to-mass ratios due to large-scale magnetic fields, making their motion analogous to that of charged test particles in a weakly charged spacetime. Furthermore, while massless charged particles are not established in the Standard Model, they serve as a useful theoretical limit to isolate the effects of charge from mass.

Our analysis is currently limited to static, spherically symmetric black holes. Real astrophysical black holes are expected to be Kerr-like. Extending this topological approach to rotating black holes (Kerr-Newman metrics) is a crucial next step, as frame-dragging effects may introduce richer topological structures. Based on the findings for rotating neutral black holes, we conjecture that the total topological charge for rotating charged black holes with flat, AdS, or dS asymptotics would still be governed by the asymptotic and horizon boundary conditions, but this requires rigorous verification.

Finally, connecting these topological invariants to observable phenomena remains an open challenge. The stability of circular orbits directly influences the fundamental frequencies of quasi-periodic oscillations (QPOs) in X-ray binaries and the characteristics of the black hole shadow. A future research direction could involve exploring whether the topological charge $W$ leaves an imprint on these observables, potentially offering a new way to probe gravity in the strong-field regime. For instance, in black hole X-ray binaries, the frequencies of QPOs are linked to the orbital and epicyclic motions of accreting material. A change in the topological number—such as from $W=-1$ to $W=0$—could indicate a transition in the number or stability of circular orbits, potentially altering the QPO frequency relationships. Similarly, the shadow of a black hole is sensitive to the structure of the photon sphere. If charged particles (or effective charged degrees of freedom in magnetized plasmas) modify the null circular orbit structure, this could in principle be reflected in the shadow diameter or subring structure.


\section*{Acknowledgement}
This work is supported by the National Natural Science Foundation of China (Grant No. 12305232), the Sichuan Science and Technology Program (Grant No. 2025ZNSFSC0830), and the Research Start-up Funding of Chengdu University of Technology (Grant No. 10912-KYQD2022-09307).

\appendix
\section{The case of V=$V_-$}\label{Appendix}
The results for $V=V_-$ are summarized in Tab.~\ref{AppendixA}, which differ from the $V=V_+$ case primarily in the sign of the charge-related terms.
\begin{table}[H]
	\centering
	\caption{For V=$V_-$, the topology of circular orbits for charged particles in spacetime.}
	\includegraphics[width=5in]{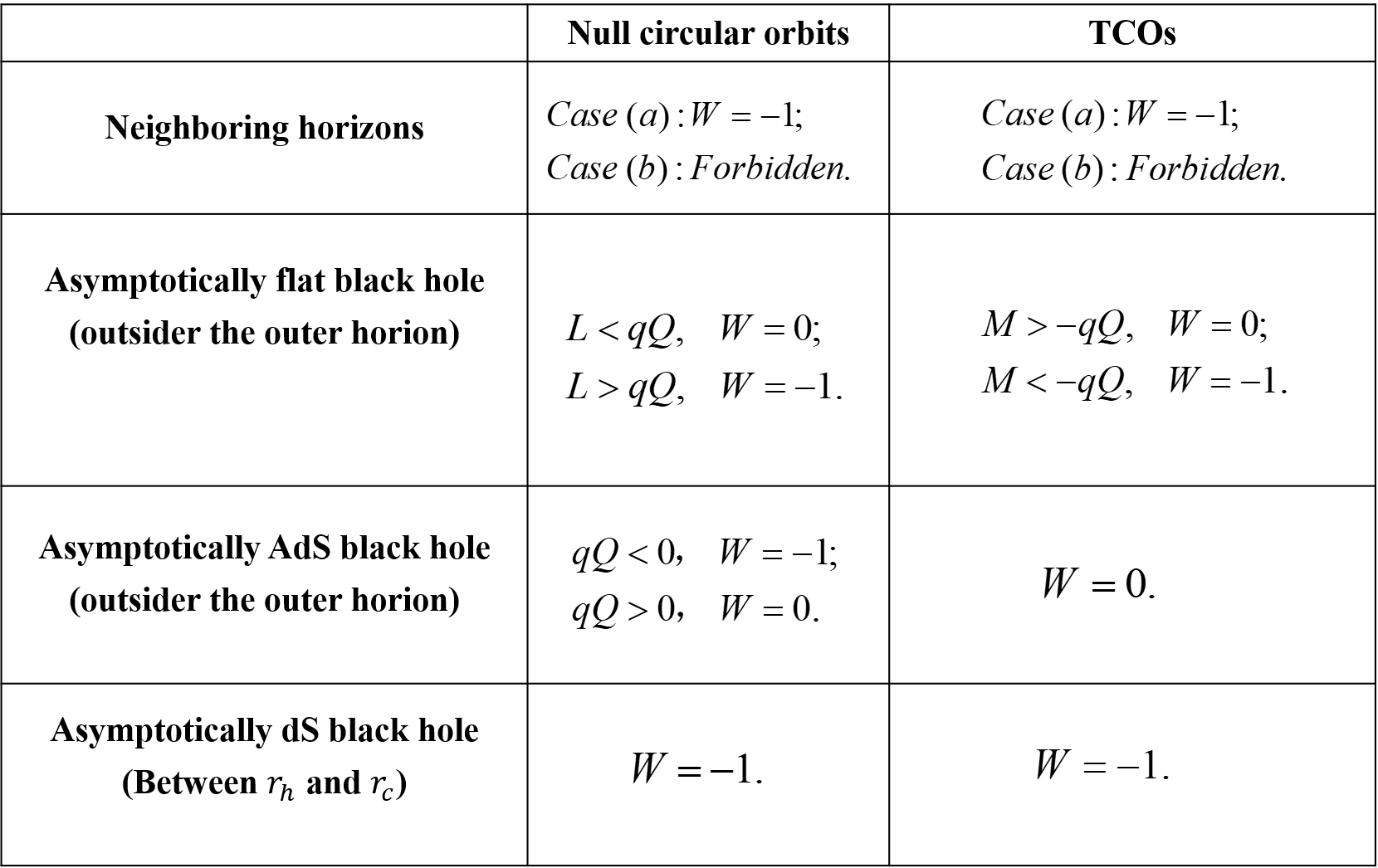}
	\label{AppendixA}
\end{table}

\end{document}